\documentclass{jpp}
\usepackage{graphicx}
\usepackage{hyperref}

\usepackage{amsfonts}
\usepackage{times}
\usepackage{graphicx}
\usepackage{natbib}
\usepackage{color}

\def\q{\qquad}

\renewcommand{\vec}[1]{\mbox{\boldmath $#1$}}
\def \d{{\rm d}}

\def \Om  {{\it \Omega}}
\def \Omin  {{\it \Omega_{\rm in}}}
\def \Omout  {{\it \Omega_{\rm out}}}
\def \rin {r_{\rm in}}
\def \Rin {R_{\rm in}}
\def \Rout {R_{\rm out}}
\def \Rey {\ensuremath{\rm{Re}}}
\def \Ha {\ensuremath{\rm{Ha}}}

\def \Pm {\ensuremath{\rm{Pm}}}
\def \Rm {\ensuremath{\rm{Rm}}}
\def \Mm {\ensuremath{\rm{Mm}}}
\def \S  {\ensuremath{\rm{S}}}

\def\beg{\begin{equation}}
\def\ende{\end{equation}}
\newcommand{\gsim}{\lower.7ex\hbox{$\;\stackrel{\textstyle>}{\sim}\;$}}
\newcommand{\lsim}{\lower.7ex\hbox{$\;\stackrel{\textstyle<}{\sim}\;$}}
\renewcommand{\vec}[1]{\mbox{\boldmath $#1$}}

\def\ara\&a{ Ann. Rev. Astronomy Astrophysics}

\shorttitle{SMRI for Couette flows of various gaps}
\shortauthor{G. R\"udiger  and M. Schultz}

\title{The gap-size influence on the excitation of  magnetorotational instability in cylindrical Couette flows
\\   {\small \today}
}

\author{G. R\"udiger\aff{1,2}
  \corresp{\email{gruediger@aip.de}}, 
\and  M. Schultz\aff{2}
 }

\affiliation{\aff{1}University of Potsdam, Institute of Physics and Astronomy, Karl-Liebknecht-Str. 24-25, 14476 Potsdam, Germany
\aff{2}Leibniz-Institut f\"ur Astrophysik Potsdam, An der Sternwarte 16, D-14482 Potsdam, Germany
}

\begin{document}

\maketitle

\begin{abstract}
The excitation conditions of the magnetorotational instability  are studied for axially unbounded Taylor-Couette flows of various gap widths between the cylinders. The cylinders are considered as made from both perfect-conducting or   insulating material and the conducting fluid with a finite but  small magnetic Prandtl number rotates with a quasi-Keplerian velocity profile. The solutions are optimized with respect to the wave number {\em and} the Reynolds number of the rotation of the  inner cylinder. For the axisymmetric modes  we find the critical Lundquist number of the applied axial magnetic field the lower the wider the gap between the cylinders. A similar result is obtained for  the induced cell structure: the wider the gap the more spherical the cells are. The  marginal rotation rate of the inner cylinder -- for fixed size of the outer cylinder --   always possesses a minimum for not too wide and not too narrow gap widths. For perfect-conducting walls the minimum lies at $\rin\simeq 0.4$ while it is at $\rin\simeq 0.5$ for insulating walls where $\rin$ is the normalized radius of the inner cylinder. The lowest magnetic field amplitudes to excite the instability are required for Taylor-Couette flows between perfect-conducting cylinders with  gaps corresponding to  $\rin\simeq 0.2$. For even wider and also for very thin gaps the needed  magnetic fields and rotation frequencies are shown to become rather huge.

Also the nonaxisymmetric modes with $|m|=1$ have been considered. Their excitation generally requires stronger magnetic fields and higher magnetic Reynolds numbers in comparison to  those for the axisymmetric modes which is primarily true for wide-gap containers with $\rin\lsim 0.3$.
\end{abstract}
\section{Introduction and motivation}
Couette flows with conducting fluids  between two rotating cylinders are favorable for an experimental realization of the various versions  of the magnetorotational instability (MRI)  for which   the applied magnetic field  is always  current-free \citep{V59,RZ01,JG01,SJ09,SJ12}.  If the field is spiral rather than axial the necessary  Reynolds numbers and Hartmann numbers are surprisingly  small  \citep{HR05,RH06} which made it possible  to investigate the corresponding helical version of MRI (HMRI) in much detail in the  {\sc Promise} experiment \citep{SG06}. The resulting instability modes are axisymmetric and they are  migrating along the rotation axis. The closely related azimuthal MRI (AMRI) appears when working  with current-free toroidal fields for which  the unstable modes are nonaxisymmetric \citep{OP96,HT10,SS14}. 

The experiments mentioned above have used one and the same container construction where the inner  radius was 50\% of the outer radius, i.e. $\rin=\Rin/\Rout=0.5$.    For this geometry the rotation shear $\mu=\Omout/\Omin$ for quasi-Keplerian rotation is  $\mu=\rin^{1.5}=0.35$. In a recent paper the Princeton group presented experimental results related to the standard version of MRI (SMRI), with a purely axial field being applied, using a  container of $\rin=0.35$ and the aspect ratio $\Gamma=H/(\Rout-\Rin)=2.1 $ \citep{WG22}. 

After previous results the absolute minimum of the magnetic Reynolds number (optimized with respect to axial wave number) necessary for the excitation of marginal instability is 21 for perfectly conducting walls and 14 for insulating walls for $\rin=0.5$ and for a shear flow with  $\mu=0.33$ \citep{RG18}. For their experiment \cite{WG22} report a rotation rate ratio of $\mu=0.19$  slightly smaller than the value 0.21 for quasi-Keplerian rotation.


The model with a  homogeneous fluid
contained between two vertically-infinite rotating cylinders is used with an applied uniform
 magnetic field parallel to the rotation axis. 
For viscous flows the most general form of the rotation law $\Om(R)$ of the
fluid is
\begin{equation}
{\Om}(R) = a+{b\over R^2},
\label{Om}
\end{equation}
where $a$ and $b$ are two constants related to the angular
velocities $\Om_{\rm{in}}$ and $\Om_{\rm{out}}$ with which the inner
and the outer cylinders are rotating and $R$ is the distance from the
rotation axis. If $R_{\rm{in}}$ and $R_{\rm{out}}$
($R_{\rm{out}}>R_{\rm{in}}$) are the radii of the two cylinders then
\begin{equation}
a={ \mu-{\rin}^2\over1-{\rin}^2}{\Om}_{\rm in}
\ \ \ \ \ {\textrm{and}} \  \ \ \  
b= R_{\rm{in}}^2 {1-\mu \over1-{\rin}^2} {\Om}_{\rm in}
\label{ab}
\end{equation}
with the geometry parameters
\begin{equation}
\mu={{\Om}_{\rm{out}}\over\Om_{\rm{in}}}  \q  {\textrm{and}} \q 
\rin={R_{\rm{ in}}\over R_{\rm{out}}}.
\label{mu}
\end{equation}
Following the Rayleigh stability criterion, 
\begin{equation}
{\d (R^2 {\Om})^2\over \d R}>0,
\label{Ray}
\end{equation}
rotation laws are hydrodynamically stable
for $a>0$, i.e. $\mu>\rin^2$. They should in particular  be stable for
resting inner cylinder, i.e. $\mu \to \infty$.

The present paper has two motivations. The first one concerns the question of the dependence of the eigenvalues on the gap width of the container. The aspect ratio $\rin$ of the cylinder radii is the only free parameter describing the geometry of the axially unbounded container. It is so far unknown how the gap width determines the critical rotation rate for given magnetic field and also the wave number of the excited instability pattern. The latter result will have  consequences for necessary vertical extension  of a possible experimental setup. 

We work with the magnetic Prandtl number 
\begin{equation}
{\rm Pm} = {\nu\over\eta},
\label{pm}
\end{equation}
 with $\nu$ as the
kinematic viscosity and $\eta$ as the magnetic diffusivity. 
The equations  are 
solved here  for the small magnetic Prandtl number 
Pm $=10^{-5}$  close to the value for liquid sodium. As known, for small magnetic Prandtl numbers the excitation conditions for the standard magnetorotational instability only depend on the microscopic magnetic diffusivity rather than the molecular viscosity, hence they do not  depend on the actual value of $\Pm$. The
 ratio of the container wall radii  is varied from  $\rin= 0.1$  to $\rin= 0.9$. We shall see that  between these values the critical magnetic Reynolds number of rotation possesses a minimum while the critical Lundquist number of the applied magnetic field linearly grows with the aspect ratio $\rin$. 

The second question  to be  attacked is  the difference of the excitation conditions for axisymmetric and nonaxisymmetric modes. Though it is widely known that the mode with the easiest excitation is the axisymmetric one  it is still important to know how much more difficult  the excitation of a nonaxisymmetric mode is.

\section{Basic equations}
The MHD equations which have to be solved are 
\beg
{\partial {\vec u} \over \partial t} + ({\vec u} \nabla){\vec u} = - {1\over \rho}
\nabla p + \nu \Delta {\vec u} +{1\over \rho} {\vec J} \times {\vec B}
\label{0}
\ende
and
\beg
{\partial {\vec B} \over \partial t}= {\rm curl} ({\vec u} \times {\vec B}) + \eta \Delta{\vec B},
\label{0.1}
\ende
with the electric current
\begin{equation}
 {\vec J}= {1\over \mu_0}{\rm curl} {\vec B}
\label{J}
\end{equation} 
  and
$
{\rm div}\ {\vec u} = {\rm div} {\vec B} = 0$.
They are considered in cylindrical geometry with 
$R$, $\phi$, and $z$ as the coordinates.  A viscous 
electric-conducting incompressible fluid between
two rotating infinite cylinders in the presence of a uniform magnetic
field  parallel to the rotation axis leads to the basic solution
$U_R=U_z=B_R=B_\phi=0,
B_z=B_0={\rm const} \ {\rm and} \  U_\phi=a R+b/R$, 
with  ${\vec U}$ as  the flow  and  ${\vec B}$ as the magnetic field. 
We are interested in the stability of
this solution. The perturbed state of the flow may be  described by
$u'_R, \; u'_\phi, \; u'_z, \; p', B'_R, \; B'_\phi, \; B'_z$
with  $p'$ as the  pressure perturbation.

In the following only  the linear stability problem will be considered.
By expansion of  the disturbances into normal modes  the solutions
of the linearized equations are of the form
\begin{eqnarray}
{\vec u}'={\vec u}(R)\ {\rm e}^{{\rm i}(m\phi+kz+\omega t)},\ \ \  \ \ \ \ \ \ \ \ \ {\vec B}'={\vec B}(R)\ {\rm e}^{{\rm i}(m\phi+kz+\omega t)}. 
\end{eqnarray}
From here on all dashes  are being omitted from the symbols of fluctuating quantities. The marginal stability line is defined  where  the imaginary part   $\Im({\omega})$
 vanishes.
We shall always use the geometrical average 
\begin{equation}
R_0=\sqrt{(R_{\rm out} - R_{\rm in})R_{\rm in}}
\label{2.4}
\end{equation}
 as the unit of length, $ \eta /R_0$ as the unit of
 velocity and $B_0 $ as the unit of the
magnetic field.
We note the rather  weak dependence of  $R_0$ on the value of $\rin$ as long as  $0.2\leq\rin \leq 0.8$. The $R_0$ only becomes small for $\rin\to 0$ or for $\rin\to 1$, i.e. for very wide or for very narrow gaps between the cylinders. Its maximum is reached for $\rin=0.5$. In order to denormalize quantities, $R_0^{-1}$  is used as  the unit of wave numbers and $\nu/R_0^2$ as  the unit of frequencies.

Using the same symbols for normalized quantities, the equations
can be written as a system of 10 equations of first order, i.e.
\begin{eqnarray}
&& \frac{\d u_R}{\d R}= -\frac{u_R}{R} - {\rm i}\frac{m}{R} u_\phi - {\rm{i}}ku_z,\\
&&\frac{\d u_\phi}{\d R} = X_2-\frac{u_\phi}{R},\\
&& \frac{\d u_z}{\d R}=X_3,\\
&&\frac{\d X_1}{\d R}=\left(\frac{m^2}{R^2}+k^2\right)u_R + 
{\rm i}(\omega+m{{\rm Re}}\,\Omega) u_R+\nonumber\\
&& \quad \quad \quad\quad\quad\quad\quad\quad\quad+ 2{\rm{i}}\frac{m}{R^2} u_\phi
-2{\rm{Re}}\,\Omega u_\phi-{\rm{i}}k{\rm{Ha}}^2 B_R,\label{dx1}\\
&& \frac{\d X_2}{\d R}= \left(\frac{m^2}{R^2}+k^2\right)u_\phi +{\rm{i}}(\omega+
m{\rm{Re}}\,\Omega) u_\phi - \nonumber\\
&& \quad\quad\quad\quad\quad\quad\quad\quad\quad- 2{\rm{i}}\frac{m}{R^2} u_R 
+ 2a {\rm{Re}} u_R -{\rm{i}}k {\rm Ha}^2 B_\phi +\nonumber\\
&& \quad\quad\quad\quad\quad\quad\quad\quad\quad + \frac{m^2}{R^2} u_\phi + k \frac{m}{R} u_z - {\rm{i}} \frac{m}{R} X_1,\\
&& \frac{\d X_3}{\d R}= \left(\frac{m^2}{R^2}+k^2\right)u_z + {\rm{i}}
(\omega+m{\rm Re}\,\Omega)u_z -\nonumber\\
&& \quad\quad\quad\quad\quad\quad\quad\quad\quad- \frac{X_3}{R} -
{\rm i}k{\rm{Ha}}^2 B_z + k \frac{m}{R} u_\phi + k^2 u_z -{\rm{i}}kX_1,\\
&&\frac{\d B_R}{\d R}= -\frac{B_R}{R} -{\rm i} \frac{m}{R} B_\phi -{\rm i}kB_z,\\
&&\frac{\d B_\phi}{\d R}= X_4 - \frac{B_\phi}{R},\\
&&\frac{\d B_z}{\d R}={\rm{i}}\left( \frac{m^2}{kR^2} +k \right)B_R
- \frac{{\rm Pm}}{k}(\omega+m{\rm Re}\,\Omega)B_R + u_R-\frac{m}{kR}X_4,\label{24}\\
&& \frac{\d X_4}{\d R}=\left( \frac{m^2}{R^2} +k^2 \right)B_\phi
+{\rm i} {\rm Pm}(\omega+m{\rm Re}\,\Omega)B_\phi-\nonumber\\
&& \quad\quad\quad\quad\quad\quad\quad\quad\quad- 2{\rm{i}}\frac{m}{R^2}B_R-{\rm i} ku_\phi
+2{\rm Pm Re}\frac{b}{R^2}B_R.
\label{sys10},
\end{eqnarray}
where $X_1$ is given by
\beg
X_1= \frac{\d u_R}{\d R} + \frac{u_R}{R} - P - {\rm Ha}^2 B_z
\label{X1}
\ende
with $P$ as the pressure fluctuation.

Here the dimensionless Reynolds number $\Rey$ and the Hartmann number 
$\Ha$ are defined as
\begin{equation}
 {\rm Re} = {R_0^2 {\Om}_{\rm in} \over \nu}\ \ \ \ \ \ \ \ \ \ \ \ \ \ \ \ \ \ \ 
{\rm Ha} = 
{R_0 B_0 \over \sqrt{
\mu_0 \rho \nu \eta}}.
\label{HA}
\end{equation}
For  given Hartmann number and magnetic Prandtl number we 
shall compute with a linear theory the critical Reynolds number of the 
rotation of the inner cylinder, also for various azimuthal mode numbers $m$. We shall see that the excitation conditions for SMRI can easily be expressed by the magnetic Reynolds number $\Rm
$ and the Lundquist number $\S$, with the definitions
\begin{equation}
 {\Rm} = {R_0^2 {\Om}_{\rm in} \over \eta}\ \ \ \ \ \ \ \ \ \ \ \ \ \ \ \ \ \ \ 
{\S} = 
{R_0 B_0 \over 
\sqrt{\mu_0 \rho}  \eta}
\label{S}
\end{equation}
without  influence of the molecular viscosity. The ratio of both quantities forms the magnetic Mach number of rotation,
\begin{equation}
 \Mm = {\Rm \over \S},
\label{Mm}
\end{equation}
which describes the strength of the rotation normalized with the applied magnetic field. The majority of  cosmic objects is characterized by  magnetic Mach numbers larger than unity (except the magnetars). We shall  discuss the run of the magnetic Mach number on the normalized   gap width between the cylinders only for  the characteristic constellation where the Reynolds number is minimal for the excitation of the instability. 

If for given size of the container and for given magnetic diffusivity the rotation of the inner cylinder is needed for which the instability is excited then the alternative dimensionless numbers
\begin{equation}
 {\Rm}_{\rm out} = {\Rout^2 {\Om}_{\rm in} \over \eta}
\ \ \ \ \ \ \ \ \ \ \ \ \ \ \ \ \ \ \ 
{\S}_{\rm out} = 
{\Rout B_0 \over 
\sqrt{\mu_0 \rho}  \eta}
\label{rmout}
\end{equation}
can be used for the  normalized rotation rate and magnetic field strength, respectively.

The below calculations were specifically carried out for the small magnetic Prandtl number $\Pm=10^{-5}$ but as we know that the obtained  eigenvalues $\Rm $ and $\S$ are also correct for even smaller $\Pm$ \citep{RS02}.
The reason is that for small $\Pm$ the critical Reynolds number runs with $1/\Pm$ so that the magnetic Reynolds number $\Rm\simeq$~const. It is clear, therefore, that for $\Pm=0$ the magnetorotational instability does not exist.
\section{Boundary conditions}
For the solution of the differential equations of 10th order a set of 10 boundary conditions is needed. 
Always no-slip conditions for the velocity on the walls
are used, i.e. 
$
u_R=u_\phi=\d u_R/\d R=0.$
 The magnetic boundary conditions depend on the electrical properties
of the walls. For perfectly conducting walls the tangential currents and the radial component of the 
magnetic field  vanish  hence
$
\d B_\phi/\d R + B_\phi/R = B_R = 0$. These   boundary conditions  hold for both 
 $R=R_{\rm in}$ and  $R=R_{\rm out}$.

For insulating  walls the magnetic boundary conditions are different at 
$R_{\rm in}$ and $R_{\rm out}$, i.e.  for $R_{\rm in}$
\begin{equation}
B_R+{\rm i} \frac{B_z}{I_m(kR)} \left(\frac{m}{kR} I_m(kR)+I_{m+1}(kR)\right)=0,
\label{72.7}
\end{equation}
and for $R=R_{\rm out}$ 
\begin{equation}
B_R+{\rm i} \frac{B_z}{K_m(kR)} \left(\frac{m}{kR} K_m(kR)-K_{m+1}(kR)\right)=0,
\label{72.8}
\end{equation}
where $I_m$ and $K_m$ are the modified Bessel
functions.
The condition for the toroidal field is 
$k R B_\phi =m B_z$ \citep{RS03}.
Neutral stability of the solutions is reached for vanishing $\Im(\omega)$. 

The homogeneous set of equations (2.6) -- (\ref{sys10})  with the boundary
conditions  included  determine the eigenvalue problem of the form 
$
{\cal L}(k, m,  {\rm Re}, {\rm Ha}, \omega)=0
$ 
for given Pm. $\cal L$ is a complex quantity, both its real part and its imaginary part 
must vanish for the critical Reynolds number.  
For nonaxisymmetric modes the real part,   ${\Re}(\omega)$, of $\omega$ describes a  
drift of the pattern along the azimuth. 
It 
is the second quantity  fixed by the complex eigenequation.    
For a fixed Hartmann number, a fixed Prandtl number and a given vertical wave 
number we also find the critical $\Rey$ of the  system. It is 
minimal for a certain wave number  defining a marginally
unstable mode. The corresponding value  $\Rey_{\rm min}$ is the minimal Reynolds number if $\Ha$ is varied and  $\Ha_{\rm min}$ is the Hartmann number for which the $\Rey_{\rm min}$ occurs. For nonaxisymmetric modes the real part of the frequency $\omega$ is the second eigenvalue   fixed by the eigenequation.    

 \section{General results}
For perfect-conducting boundary conditions the resulting curves of marginal stability are given in Fig. \ref{fig1a} and for vacuum boundary conditions in Fig. \ref{fig1b}. In both plots the left graph presents the magnetic Reynolds numbers (optimized by the wave number) and the right one gives the resulting wave numbers, in both cases as functions of the Lundquist number. The dashed lines belong to the nonaxisymmetric modes with $|m|=1$. Generally, the latter require higher values $\Rm_{\rm min}$ than the axisymmetric  modes. The curves for $m=0$ exhibit their typical shape: They are rather steep for  $\S<\S_{\rm min}$ and they are much flatter for  $\S>\S_{\rm min}$ where  $\S_{\rm min}$ is the   minimum of the function $\Rm=\Rm(\S)$. For $\S=$O(1) the influence of the microscopic diffusion stops the existence of the SMRI while for much larger $\S$ this existence is limited by too strong external magnetic fields.

The curves for the axisymmetric modes only possess a lower limit of the Reynolds number while the curves for the nonaxisymmetric modes always possess a lower and an upper limit of the (normalized)  rotation frequencies. The nonaxisymmetric modes are stabilized by too fast differential rotation \citep{Rae86,RG18}.

\subsection{Medium gaps}
\begin{figure}
  \centerline{
 \includegraphics[width=0.53\textwidth]{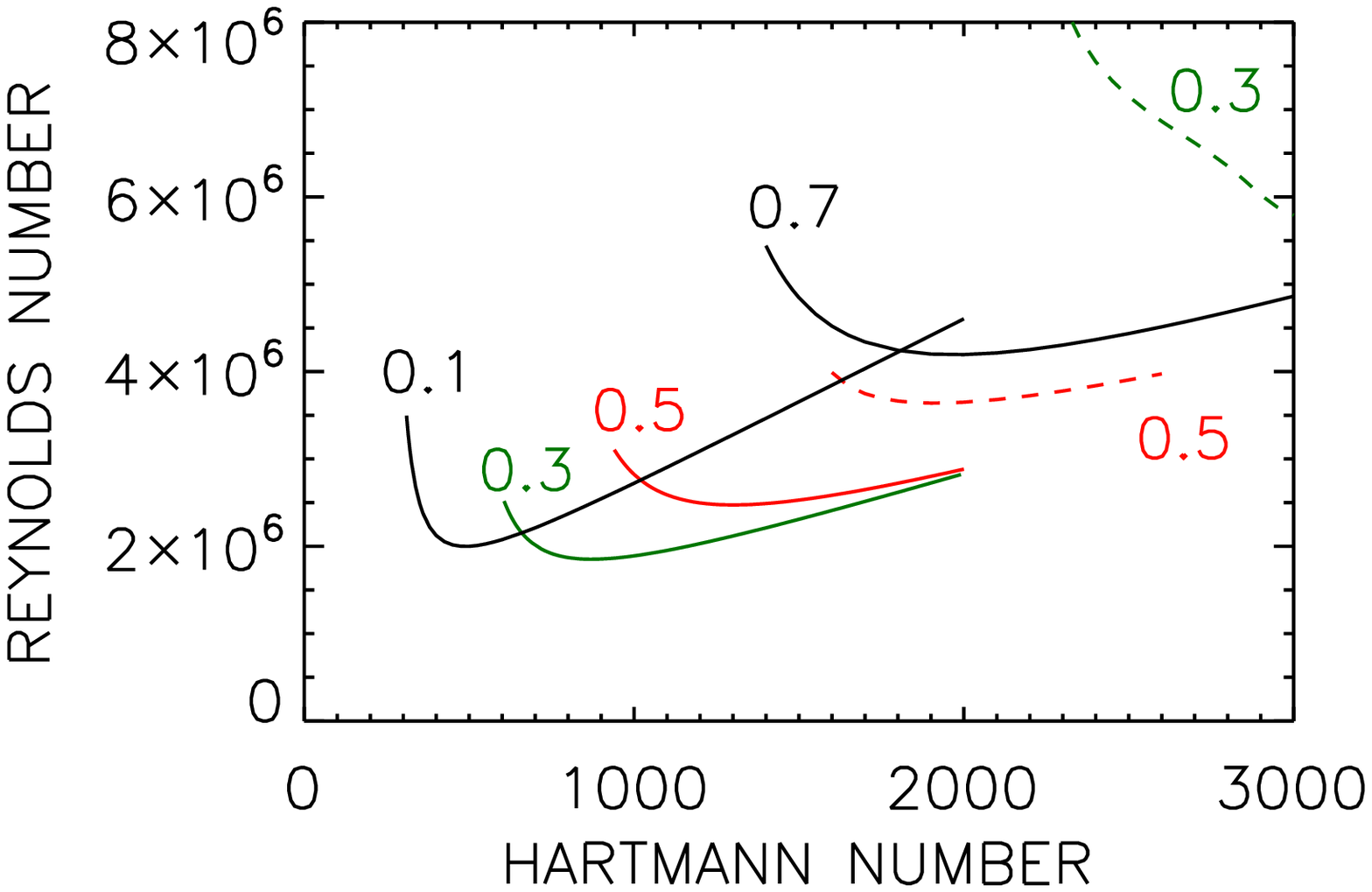}
 \includegraphics[width=0.53\textwidth]{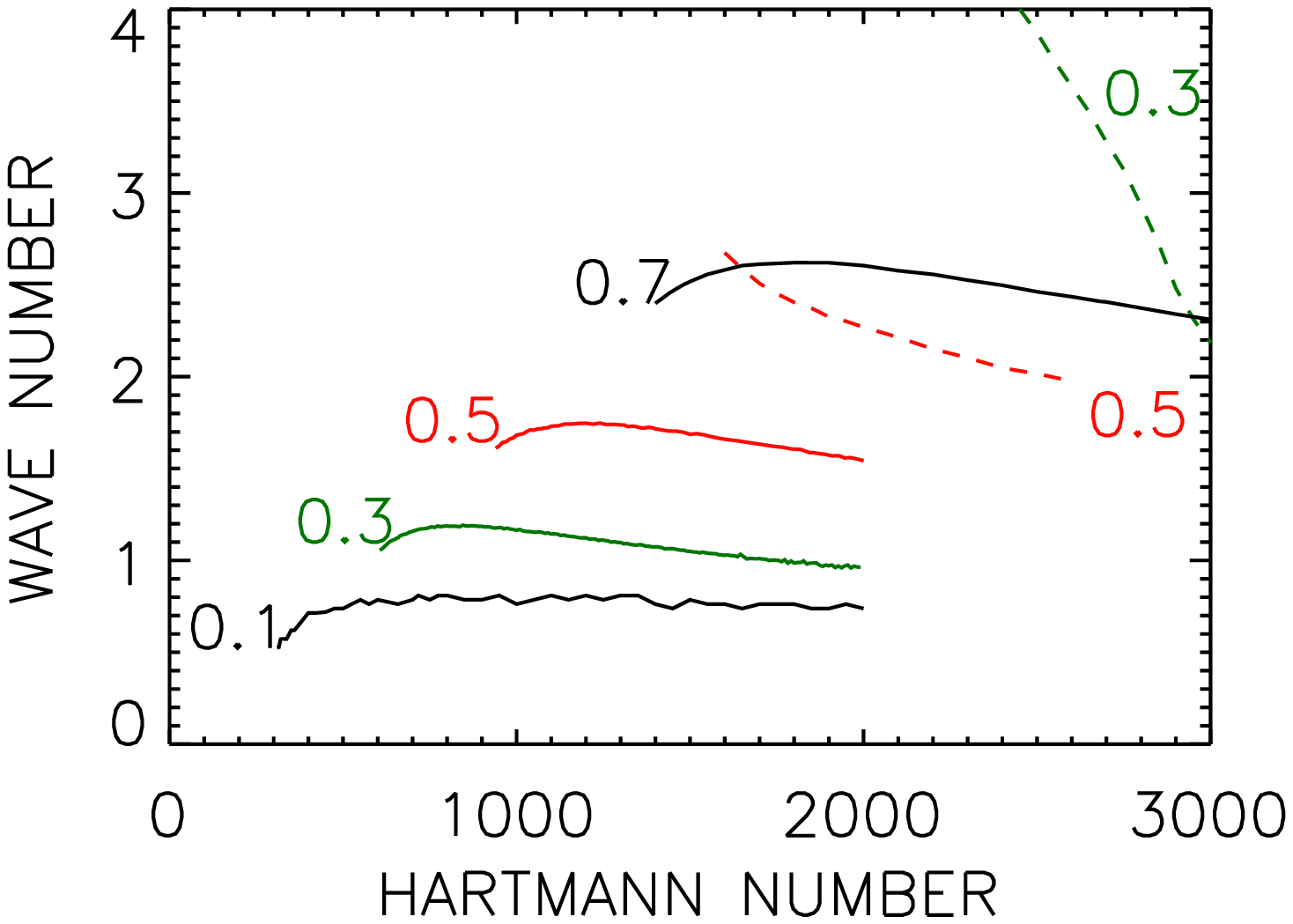}
 }
  \caption{Left: Stability maps for $m=0$ (solid lines): $\rin=0.1$, $\rin=0.2$, $\rin=0.3$ (green),  $\rin=0.4$,  $\rin=0.5$ (red),  $\rin=0.6$,  $\rin=0.7$  $\rin=0.8$.  For   $m=1$ (dashed lines):  $\rin=0.3$ (green),  $\rin=0.4$,  $\rin=0.5$ (red),  $\rin=0.6$. Right: The corresponding axial wave numbers. Quasi-Keplerian differential rotation, $\Pm=10^{-5}$, perfect-conducting cylinder material.} 
\label{fig1a}
\end{figure}

The minimal $\Rm_{\rm min}$ in Fig.  \ref{fig1a}  for medium gap widths approximately behave according to  $(1-\rin)\Rm_{\rm min}\simeq 13$ (see Table \ref{tab1}). This relation implies that the minimal rotation rates of the inner cylinder for the considered gaps  behave like 
$\Om_{\rm in}\propto 1/(1-\rin)$, i.e. in containers with  wider gaps the instability is easier to excite. For very wide gaps, however, both the magnetic Reynolds number as well as the needed rotation rate of the inner cylinder grow to very large values.
Minimal rotation rates are  only possible for experiments with medium $\rin$.



\begin{figure}
  \centerline{
 \includegraphics[width=0.53\textwidth]{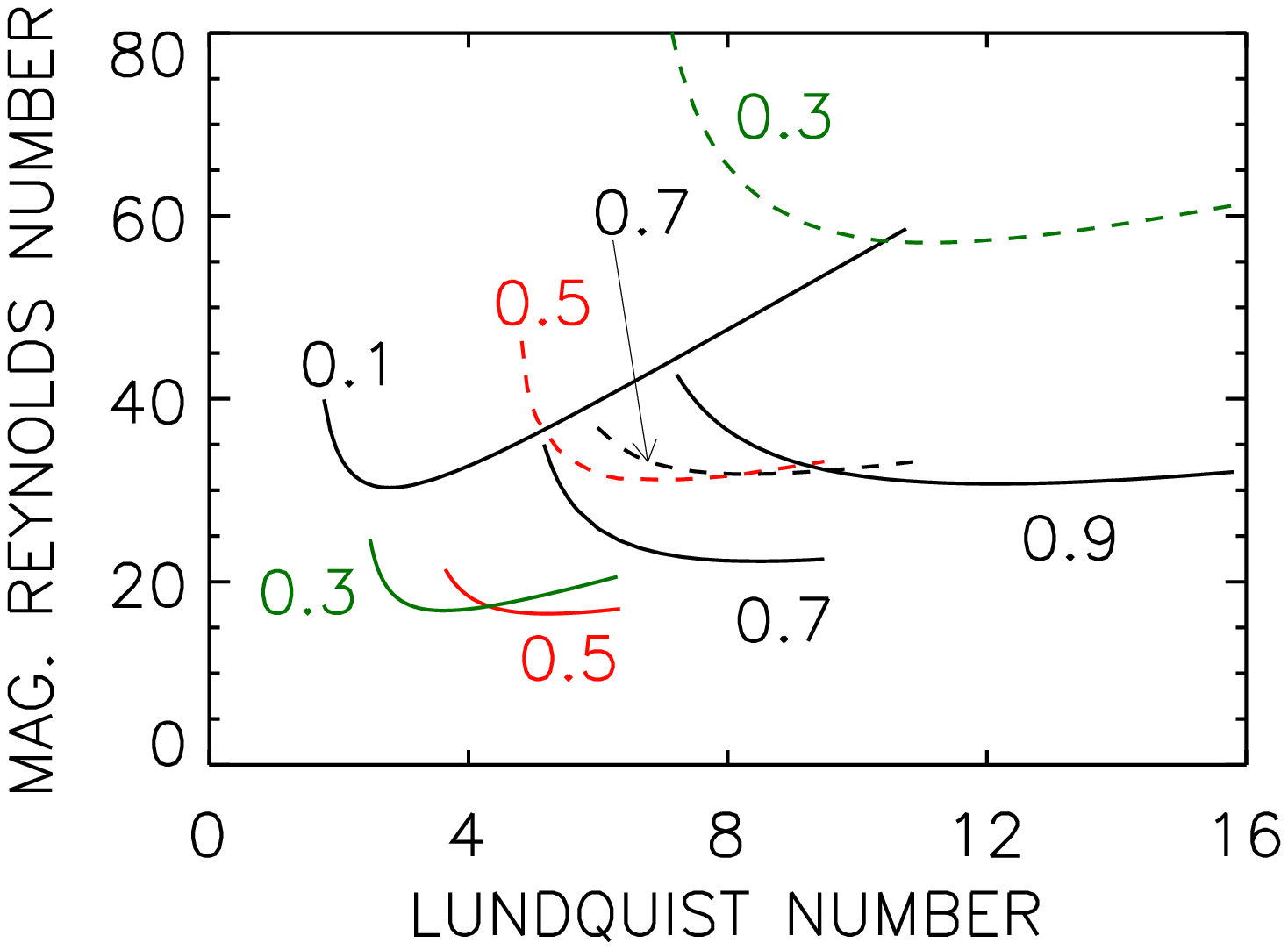}
 \includegraphics[width=0.53\textwidth]{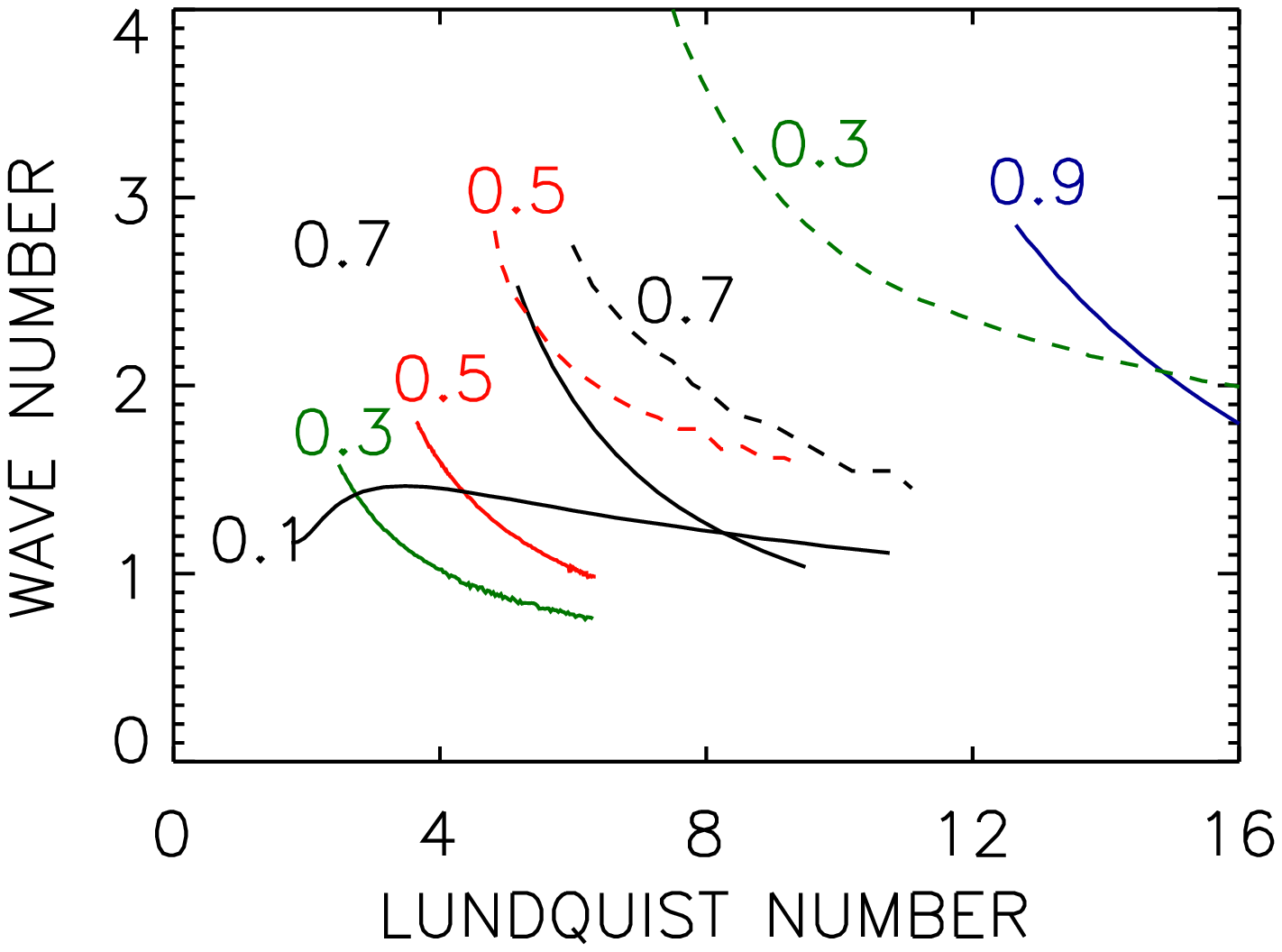}
 }
  \caption{The unstable modes for containers with insulating cylinders. Left: stability maps for $m=0$ (solid lines): $\rin=0.1$,   $\rin=0.3$ (green),   $\rin=0.5$ (red),  $\rin=0.7$,  $\rin=0.9$ (blue). 
 For   $m=1$ (dashed lines):  $\rin=0.3$ (green),  $\rin=0.4$,  $\rin=0.5$ (red),  $\rin=0.7$. 
Right: the corresponding axial wave numbers. Quasi-Keplerian differential rotation,  $\Pm=10^{-5}$.} 
\label{fig1b}
\end{figure}

In Table \ref{tab1} the characteristic eigenvalues  $\Rm_{\rm min}$ and  $\S_{\rm min}$ have been collected for the minima of the curves in Figs. \ref{fig1a} and \ref{fig1b} which we shall call the characteristic values. Obviously, the influences of the boundary conditions are only small for the containers with large gaps. One finds the $\Rm_{\rm min}$ with vacuum boundary condition as always smaller than for perfect-conductor conditions (except for very thin gaps)  but the  $\S_{\rm min}$ are always larger (see also Fig. \ref{fig3}). Consequently, the corresponding magnetic Mach numbers are  much larger for cylinders made from perfect conductors. For such containers  the axial wave numbers are always larger than for insulating walls. 

With our normalizations the vertical extent $\delta z$ of {\em one}  cell of the instability pattern, normalized by the gap width $D=R_{\rm out} - R_{\rm in}$ between the cylinders, is given by  
\beg
\zeta={\delta z \over D} = {\pi \over k}
\sqrt{{\rin \over 1-\rin}}.
\label{delz}
\ende
Flat cells are described by $\zeta<1$ while axially elongated cells possess $\zeta>1$. For $\rin=0.5$ it is simply $\zeta = {\pi}/{ k}$ so that for $k\simeq \pi$ the cell is almost circular. The examples given in Table \ref{tab1}, however, show that the wave numbers do not reach the value of $\pi$ for medium $\rin$ hence the cells are always elongated in the axial direction. This is true for models with both sorts of boundary conditions. For perfect-conducting cylinder walls the  $\zeta>1$ hardly varies with the $\rin$ -- the cells are always oblong. Evidently, such cells are not suitable to provide the angular momentum transport in accretion disks or galaxies.

For perfect-conducting cylinders it follows $\zeta\simeq 1.8$ for almost all $\rin$ while it can become significantly  larger  for insulating material.  The minimum axial extend of a container probing MRI pattern is thus $H\simeq 1.8 D$. The aspect ratio $H/D$ of the Princeton experiment is 2.1.  It happens that  the numerical $\zeta$-values differ by almost a factor of two for models with the same geometry but with different boundary conditions


\begin{table}
\caption{The coordinates of the minima of the profiles in Figs. \ref{fig1a} and \ref{fig1b} (both left)  for several radii the   inner cylinder for different  boundary conditions (left: perfect conduction; right:  vacuum). $m=0$,  $\Pm=10^{-5}$. All models are for quasi-Keplerian rotation law. 
 }
\label{tab1}
\centerline{
\hbox{
\begin{tabular}{lcccccc}
\hline\hline
\\
$\rin$ & $\mu$ &$\Rm_{\rm min}$ &   $ \S_{\rm min}$&$k$&$\zeta$&$\Mm$\\ \\
\hline
&&&&\\
0.1 &0.031  &20.0  & 1.58 &0.7&1.4&12.7\\
0.2 &0.089  &17.8  &{ 2.05}  &1.0&1.6&8.81 \\
0.3 & 0.16 & 18.5 & 2.76 &1.2&1.7&6.93 \\
0.4 &0.25& {20.8} &3.38&1.4&1.8&6.15\\ 
 0.5& 0.35 & 24.7  &4.11&1.7&1.8&6.00 \\
 0.6 &0.46& 31.1 &5.00 &2.1&1.8&6.22\\
0.7 &0.59&42.0  &6.32 &2.6&1.8&6.64\\
0.8 &0.72&64.1  &8.22 &3.4&1.8&7.79 \\
0.9 &0.85& 131 &12.6 &5.1&1.8&10.4\\
0.95&0.93 &265&17.6&7.3&1.9&15.0\\
  \hline
\end{tabular}
\hskip2cm
\begin{tabular}{lcccccc}
\hline\hline
\\
$\rin$ & $\mu$ &$\Rm_{\rm min}$ &   $ \S_{\rm min}$&$k$& $\zeta$& $\Mm$\\ \\
\hline
&&&&\\
0.1 & 0.031 &30.3  &2.78  &1.4&0.7&10.9\\
0.2 &0.089  & 20.8 & { 3.04} &1.1&1.4&6.84\\
0.3 & 0.16 & 16.9 & 3.62&1.1 &1.9&4.67 \\
0.4 &0.25& 16.0 &4.82&1.1&2.3&3.32\\ 
 0.5& 0.35 & { 16.5}  &5.21&1.2&2.6& 3.17\\
 0.6 &0.46& 18.3 &6.47&1.2 &3.2&2.83\\
0.7 &0.59&22.3  &8.54 &1.2&4.0&2.61\\
0.8 &0.72&30.7  &12.0& 1.2&5.2&2.55\\
0.9 &0.85&56.6  &22.1 &1.0&9.4&2.56\\
0.95&0.93&109&39.8&0.92&14.9&2.74\\
 \hline
\end{tabular}
}}
\end{table}
\subsection{Extremal gaps, thin-shell approximation}
We note that  for very wide gaps with $\rin\simeq 0.1$ the characteristic  Lundquist numbers  $ \S_{\rm min}$ -- for which the associated Reynolds number is minimal --  are  reduced to values of order unity while for very narrow gaps with $\rin\simeq 0.95$ the Lundquist numbers are   maximal. In both limits the Reynolds numbers  $ \Rm_{\rm min}$ possess enlarged values.  As also the linear dimension $R_0$ becomes small for small and/or large  $\rin$ the  necessary inner  rotation rates become very large excluding  the applicability of  containers with very wide and/or very narrow gaps between the cylinders for experiments. One finds that containers with $0.3 \lsim \rin \lsim 0.6$ require the least rotation rates for excitation of  standard magnetorotational instability. On the other hand,
for both sorts of boundary conditions  the models with $\rin \simeq 0.2$ require the weakest magnetic fields.

Table \ref{tab1} also gives the the results for very thin gaps between the cylinders up to $\rin=0.95$  \citep{DO60,DO62}. The most striking difference due to the choice of the boundary conditions is here the numerical value of the calculated magnetic Mach number (last column). For $\rin\to 1$ the characteristic Reynolds number grows to larger and larger values. We did not find a maximum of $ \Rm_{\rm min}$ for $\rin\to 1$. As the $R_0^2$ runs with $d=1-\rin$ for $d\to 0$ we find  $\Omin\to \infty$ in this limit. It  should  thus not be possible to work with a thin-shell approximation \citep{E58} for numerical or experimental realizations of the standard MRI in TC flows. 


\section{The nonaxisymmetric modes}
The excitation  of nonaxisymmetric modes requires faster rotation and stronger magnetic fields than the excitation of the axisymmetric modes (Fig. \ref{fig1a}, Fig. \ref{fig1b} , dashed lines). For  medium  $\rin$ of about 0.5 (red) the lines of marginal stability do hardly depend on the radius of the inner cylinder. As also the geometric radius $R_0$ is almost constant for different $\rin$  the rotation frequencies and magnetic field amplitudes needed to excite  nonaxisymmetric modes are almost the same for such values of $\rin$. However, for weak magnetic fields with $\S<\S_{\rm min}$ the Reynolds numbers for excitation of the $m=1$ modes are  {\em much} higher than those for the excitation of the axisymmetric modes with $m=0$. We note that the curves for the weak-field branch with $\S<\S_{\rm min}$ become very steep so that the excitation of nonaxisymmetric modes requires very rapid rotation.

For  wide gaps between the cylinders ($\rin=0.3$, green) one finds  that the lowest Reynolds number belongs to much higher Lundquist numbers than for $\rin\approx 0.5$. In addition,  the curves for weak Lundquist numbers are much steeper than the curve for the corresponding axisymmetric mode. Hence, for wide gaps with $\rin\simeq 0.3$ it is almost impossible to excite the axisymmetric and the nonaxisymmetric mode simultaneously by experiments with Lundquist numbers not much higher than unity (see \cite{WGE22}). 
\begin{figure}
  \centerline{
  \includegraphics[width=0.53\textwidth]{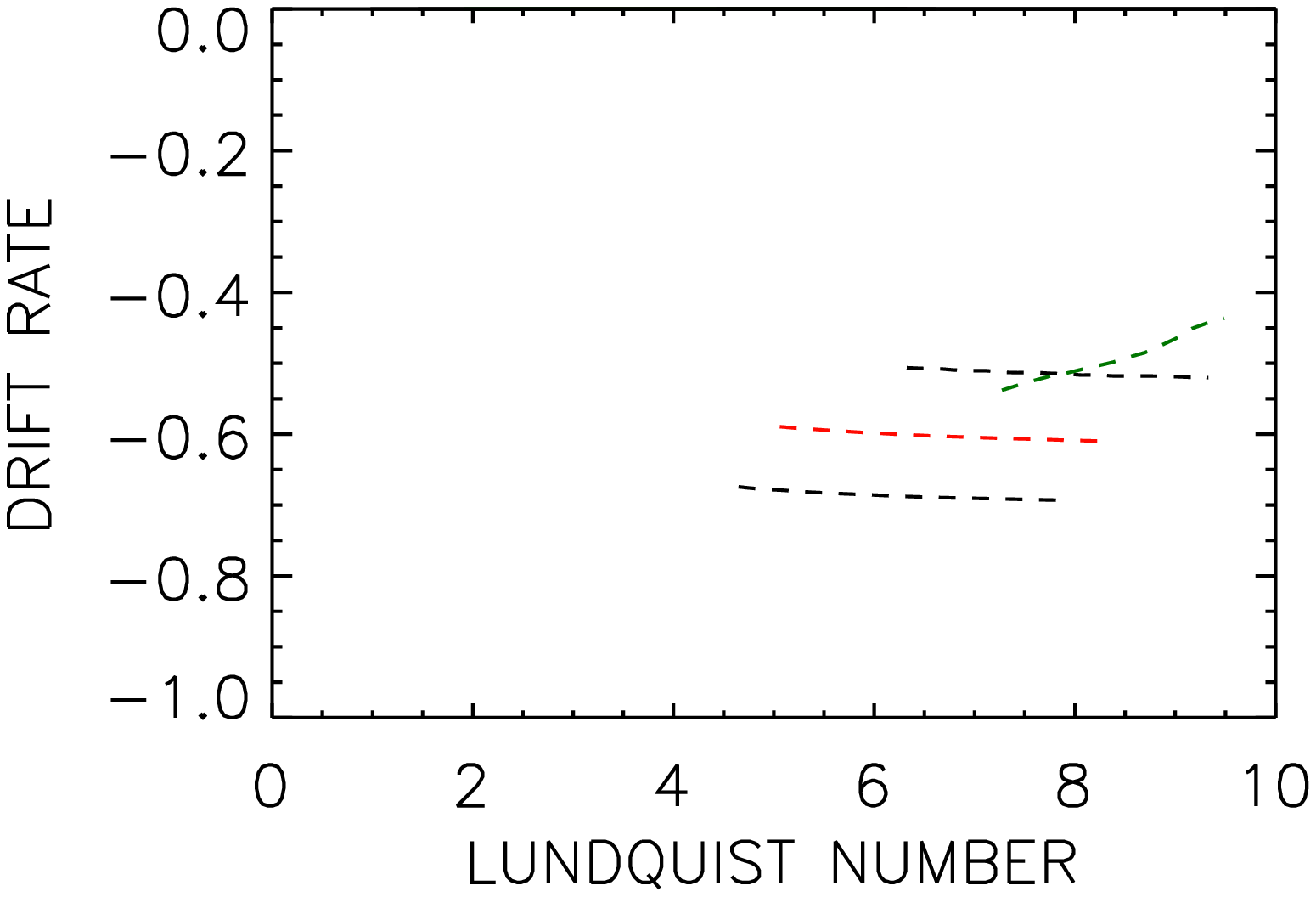}
  \includegraphics[width=0.53\textwidth]{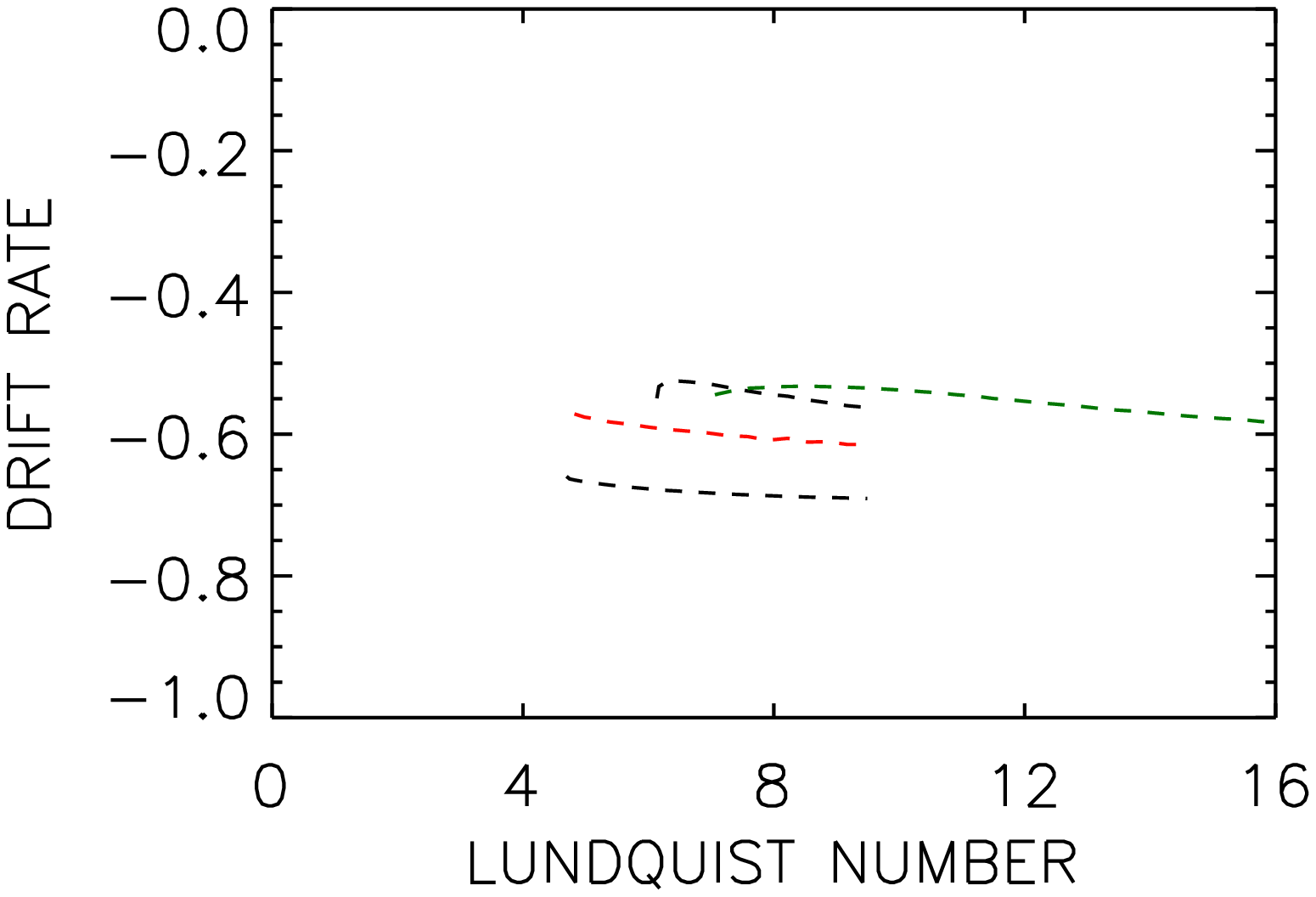}
 }
  \caption{Drift rates $\omega_{\rm dr}=\Re(\omega)/\Omin$ of the nonaxisymmetric modes   $m=1$ (dashed lines):  $\rin=0.3$ (green),  $\rin=0.4$,  $\rin=0.5$ (red),  $\rin=0.6$.  Left: perfect-conducting cylinder material, Right: insulating cylinder material.  Quasi-Keplerian differential rotation, $\Pm=10^{-5}$.} 
\label{fig1c}
\end{figure}

The nonaxisymmetric modes are drifting in azimuthal direction. The drift rates, $\omega_{\rm dr}=\Re(\omega)/\Omin$, are given in Fig. \ref{fig1c}. According to the relation
\beg
{{\partial \phi/ \partial t}\over \Omin} =  - {\omega_{\rm dr}\over m},
\label{drift}
\ende
 the negative $\omega_{\rm dr}$ plotted in the figures indicate  a migration of the patterns in the direction of the global rotation - for both sorts of boundary conditions. In all cases the azimuthal migration is faster then the rotation of the outer cylinder ($\omega_{\rm dr}>\mu$).
\section{Marginal stability  for weak fields with $\S<\S_{\rm min}$}
\begin{figure}
  \centerline{
  \includegraphics[width=0.48\textwidth]{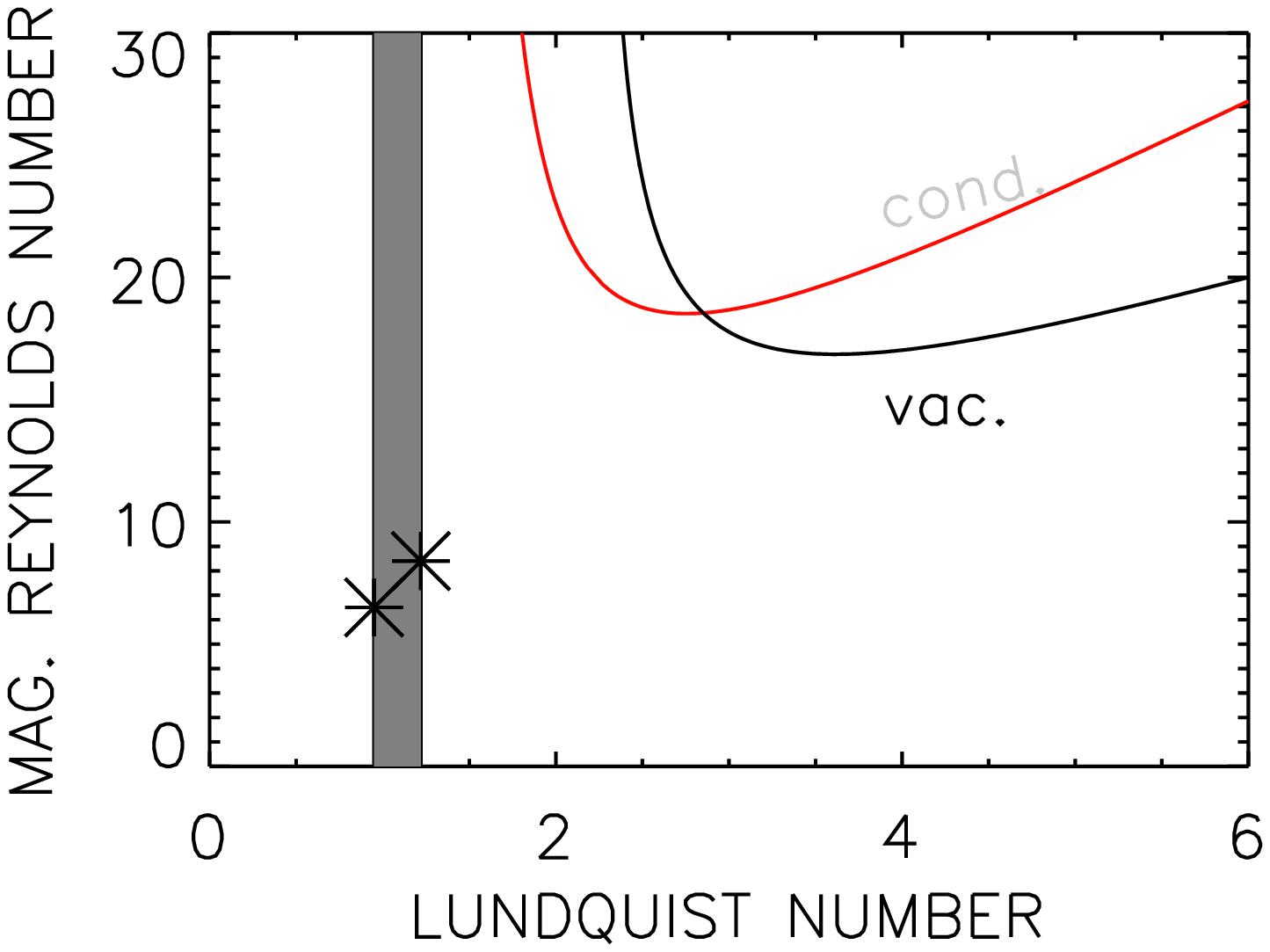}
  \includegraphics[width=0.48\textwidth]{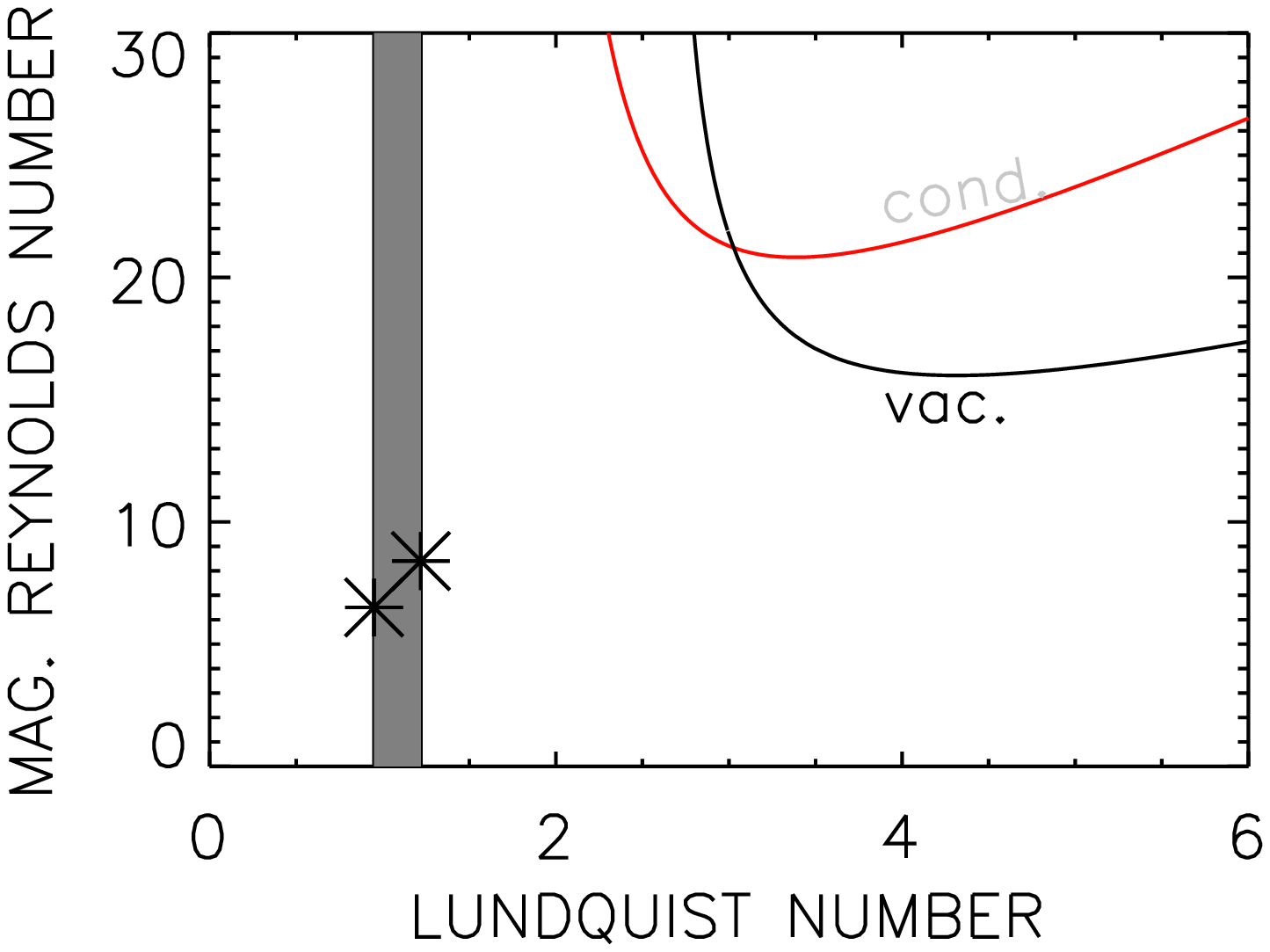}
 }
  \caption{Stability map of  the axisymmetric modes  for $\rin=0.3$ (left) and $\rin=0.4$ (right) in relation to  experimental results of \citep{WG22}. Containers with perfect-conducting cylinders (red) and with insulating cylinders (black). The vertical  lines mark the Lundquist numbers  after (\ref{S}) for the Princeton MRI-experiments represented by the  lowest and the highest solid circles in their Fig. 2a while the asterisks belong the used magnetic Reynolds numbers. The vertical lines correspond to  applied magnetic fields of 2150 G and 2750 G, resp. With the planned  {\sc Dresdyn} sodium experiment the same magnetic field  amplitudes belong to   Lundquist numbers exceeding $ 15$.} 
\label{fig3}
\end{figure}

It is  possible to operate with weak magnetic fields, i.e. with
$\S<\S_{\rm min}$. Then, however, the  Reynolds numbers necessary for excitation  basically grow and the unstable wave numbers become smaller, i.e. the cells become longer. Figures \ref{fig3} demonstrate these weak-field solutions for $\rin=0.3$ and $\rin=0.4$ where the vertical lines belong to the magnetic fields ($\S=0.96$ for $B=2150$ G and $\S=1.22$ for $B=2750$ G) reported for  gallium SMRI experiments by \cite{WG22}.
We have also to  note the  influence of the boundary conditions: for insulating cylinders the curves are so steep  that  the actual Reynolds numbers exceed the $\Rm_{\rm min}$ by more than an order of   magnitude. The effect is reduced for perfect-conducting cylinder material but still the enhancement of the critical Reynolds number is   by a factor of three relative to $\Rm_{\rm min}$. This is even a minimum value as the real cylinders are by far not perfect-conducting. For galinstan as the fluid and stainless steel as the (outer) cylinder material the ratio 
\beg
\hat \sigma = {\sigma_{\rm cyl}\over \sigma_{\rm fl}}
\label{hatsigma}
\ende
of the electric conductivities of the cylinders and the fluid is  0.47 which does neither well approach perfect-conduction nor vacuum boundary conditions. For sodium experiments one finds $\hat\sigma\simeq 0.16$, hence the vacuum boundary conditions might provide the appropriate description. \cite{RS18} derived  the form of the boundary conditions with finite values of (\ref{hatsigma}) and have shown that for $\hat\sigma$ of order unity the resulting eigenvalues can approximately  be interpolated between the values for  $\hat\sigma=0$ (insulating boundaries)  and  $\hat\sigma=\infty$ (perfect-conducting boundaries).

Detailed consequences of the application of weak magnetic fields for the excitation of the axisymmetric mode are shown in Fig. \ref{fig3}. The Lundquist number (in our definition, see (\ref{S})) is marked by  vertical lines for applied magnetic fields of 2150 G and 2750 G for a container with  $\Rout=20.3$ cm, $\rin=0.35$ filled with  galinstan ($\rho=6.4$ g cm$^{-3}$ and $\eta=2428$ cm$^2$ s$^{-1}$). The magnetic Prandtl number of galinstan is $1.4\times 10^{-6}$. These numbers   correspond to the   gallium experiment by  \cite{WG22}. The maximally possible uniform magnetic field in this experiment  given as  4800 G corresponds to a Lundquist number after Eq. (\ref{S}) of $\S=2.1$. One finds with Fig. \ref{fig3} that for such fields  the minimum magnetic Reynolds number for marginal stability must exceed 20, corresponding to a minimum rotation rate with  $Rm=10.8$ in the notation of \cite{WG22}. 
The left vertical  lines in Fig. \ref{fig3} represent the alleged SMRI realization  for $B_0=0.2$ and $Rm=3.4$ (lower asterisk)
 provided by  their Fig. 2a. The experiment with the fastest rotation ($Rm=4.5$, upper asterisk) belongs to a Lundquist number  of $\S=1.22$ (right vertical lines).

If the typical parameters  of the  experiment with almost Keplerian flow  and with maximal $Rm=4.5$
(in their notation) are transformed to our definitions (\ref{S}), one obtains maximal Reynolds numbers of  $\Rm=8.4$ (right asterisks in the plots) 
 which   does { not} reach, however, the values required for  marginal instability at {\rm any} Lundquist number.  After Fig. \ref{fig3}  this deficit is particularly drastic for vacuum boundary conditions. It is thus  risky for experiments to work with Lundquist numbers  which are basically smaller  than the  $\S_{\rm min}$ given in the Table \ref{tab1}.

Our results comply with the finding of \cite{GJ02} in their Figs. 1 and  2, that   containers with insulating walls  of $\rin=0.33$ including a quasi-Keplerian flow do not allow  the excitation of SMRI with an applied field of less than 2750 G. Even with perfect-conducting cylinders the curves are so steep for
$\S<\S_{\rm min}$ that the needed rotation rates of the  cylinders are  unrealistically high.
\section{Flat cells}
As already demonstrated in Table \ref{tab1} the cell structure for the axisymmetric modes is nearly spherical for wide gaps and rather elongated in axial direction for narrow gaps.
It is thus basically unclear whether the standard MRI is also able to produce flat cells with $\zeta \ll 1$ which are necessary to appear in experiments with a flat container ($H<D$) and/or in flat cosmical objects such as accretion disks and galaxies. For the latter, however, the magnetic Mach number of rotation does not exceed values of 10 as it is the case for the solutions given in Table \ref{tab1}.  

We have to probe, therefore,  whether eigensolutions exist for finite $\Rm$ and $\S$ when, for example, $\zeta=0.1$ is required. From (\ref{delz}) one obtains for (say)  $\rin=0.1$ that solutions with $k=10.5$  are matched for  $\zeta=0.1$ and $k=21$ for  $\zeta=0.05$.
\begin{figure}
  \centerline{
  \includegraphics[width=0.48\textwidth]{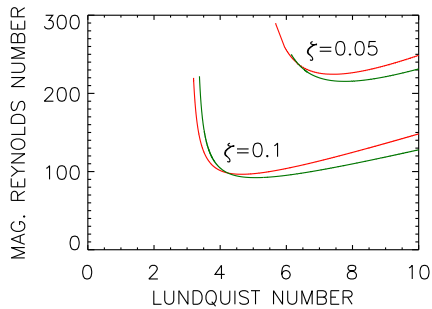}
 } 
  \caption{Lines of marginal stability for $\rin=0.1$ and a fixed axial wave number of $k=10.5$ for $\zeta=0.1$ and $k=21$ for $\zeta=0.05$. Perfect-conducting cylinders (red), insulating cylinders (green). The magnetic Mach numbers of the solutions grow for growing flatness. The influence of the boundary conditions is rather weak. $m=0$.} 
\label{fig4}
\end{figure}

Figure \ref{fig4} shows the results. Indeed, the envisaged flat cells exist in the axially unbounded container and even for similar  Lundquist numbers as for the solutions with the lowest Reynolds numbers and the elongated cells. The actual Reynolds numbers for flat cell structures, however, are much higher than before. Hence, the magnetic Mach numbers for flat cells are also higher than for the elongated cells summarized in Table \ref{tab1}. They are  larger than 10 and grow for growing flatness. This result complies with that of a global model of a flat galaxy  for magnetic Prandtl number $\Pm\geq 1$ \citep{KR04}. The more flat the cells, the stronger  the dissipation and the harder the differential rotation must work to excite the instability while the magnetic field needed for the rotation minimum  remains unchanged.
We also learn from Fig. \ref{fig4} that the influence of the actual  boundary condition is remarkably weak.

\section{Discussion}
\begin{figure}
  \centerline{
  \includegraphics[width=0.53\textwidth]{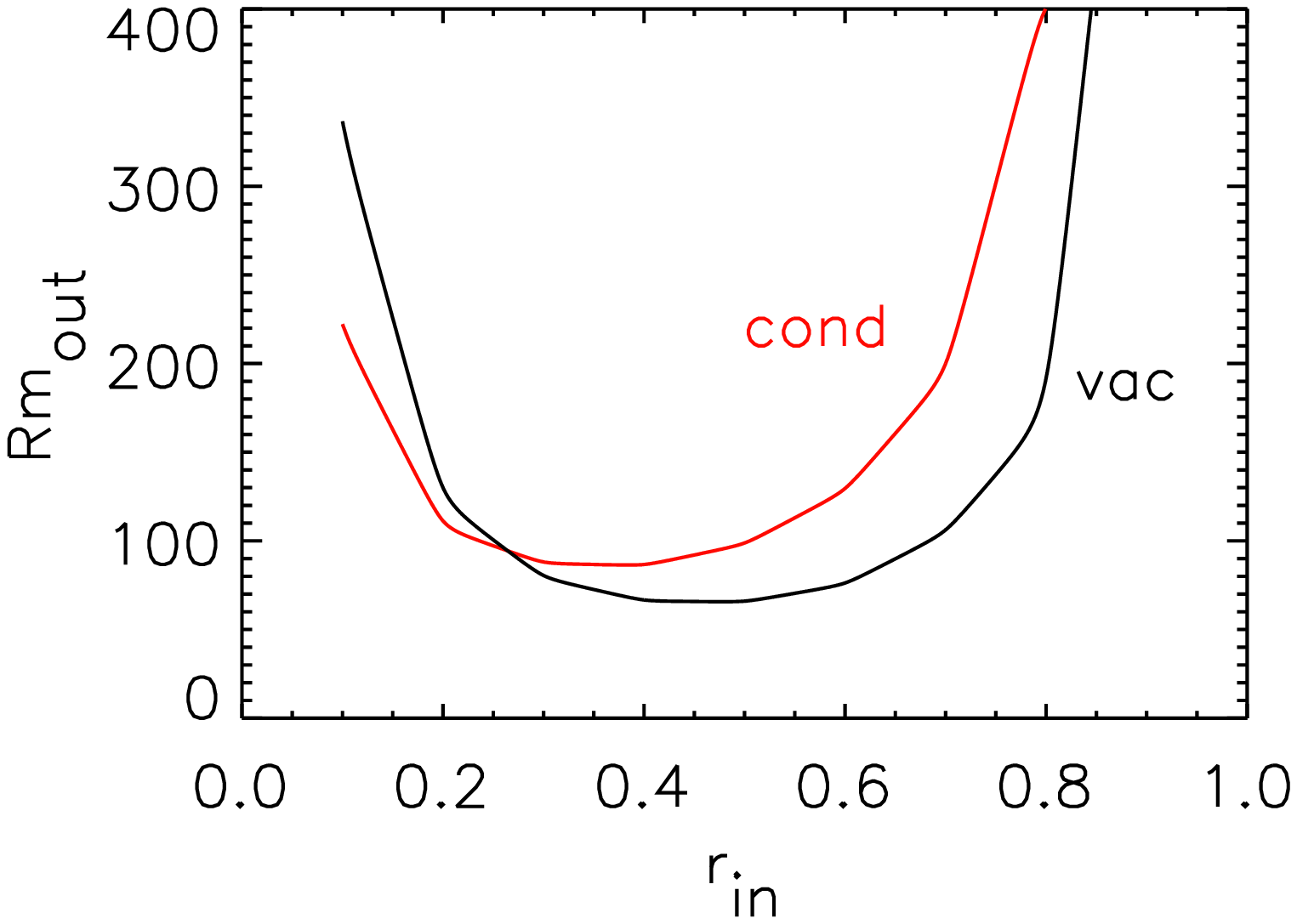}
 \includegraphics[width=0.53\textwidth]{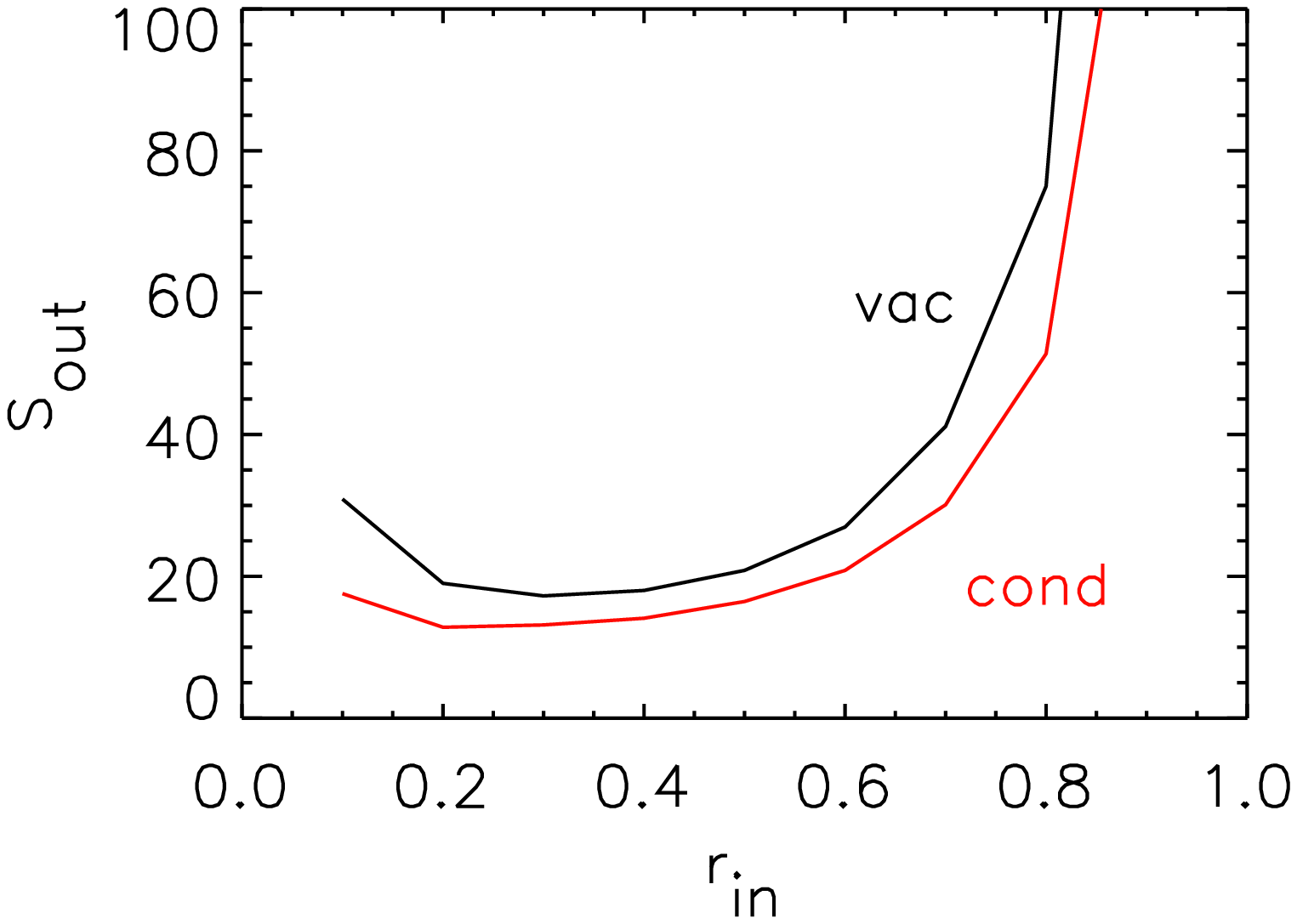}
 }
  \caption{The magnetic Reynolds number (left) and the Lundquist  number (right) after the definitions  (\ref{rmout}) representing the  normalized inner rotation rate and the magnetic field amplitude  needed for excitation vs. the inner cylinder position. Quasi-Keplerian differential rotation, $\Pm=10^{-5}$, perfect-conducting cylinder material (red), insulating cylinders (black). The numbers are taken from Table \ref{tab1}.} 
\label{Rmout}
\end{figure}

The influence of the position $\rin$ of the inner cylinder of Taylor-Couette flows on the excitation of the magnetorotational instability has been studied. To demonstrate the results we shall switch to the representations  (\ref{rmout}) of the magnetic Reynolds number and the Lundquist number which for given outer cylinder size $\Rout$ form minimal normalized inner rotation rates and magnetic field amplitudes needed for excitation of the instability. These quantities are plotted as function of $\rin$ by Fig. \ref{Rmout} for both sorts of boundary conditions. The main result is that too narrow or too wide gaps would  require very high rotation rates or very strong magnetic fields. For $0.3\lsim \rin \lsim 0.6$ the dependence of the eigenvalues on $\rin$ is rather weak. The minima for conducting cylinders are at $\rin\simeq 0.4$ for the rotation rate and at $\rin\simeq 0.2 $ for the magnetic field. We also note that the influence of the boundary conditions is opposite for rotation and field. For vacuum conditions the needed rotation rates are mostly lower than for perfect-conduction conditions but  the needed magnetic fields are higher for insulating cylinders. 

The experiment with the fastest rotation ($Rm=4.5$) by \cite{WG22} corresponds to $\Rm_{\rm out}=37$ which is certainly below the  curves in   Fig. \ref{Rmout} (left). A similar situation holds with respect to the magnetic field: the given maximal possible field of 4800 G provides a Lundquist number of $\S_{\rm out}=4.4$ which again does not reach the minimum value in the right plot of Fig. \ref{Rmout}. Compared with our calculations the reported experiments are subcritical. There is no value of $\rin$ for which the considered rotation rates and/or magnetic fields are supercritical by a large margin. For a correct interpretation of the obtained results the modification of the angular velocity profiles and the specific role of the Ekman-Hartmann layers close to the copper lids must be understood in much more details.

\acknowledgments{
Frank Stefani (Dresden-Rossendorf) is acknowledged for  discussions of the presented problem and a critical reading of the manuscript.}

\bibliographystyle{jpp}
\bibliography{superamri}
\end{document}